\documentclass[useAMS,usenatbib]{mn2e}
\usepackage{graphicx}
\usepackage[usenames,dvipsnames]{color}
\usepackage{tabularx,booktabs,ragged2e}
\usepackage{hyperref}
\usepackage{pifont}

\newcolumntype{L}{>{\RaggedRight\arraybackslash}X}

\title[A polarized view of NGC~1068]
      {A complete disclosure of the hidden type-1 AGN in NGC~1068 thanks to 52 years of broadband polarimetric observation}
\author[Marin~F.]
      {F.~Marin\thanks{E-mail: frederic.marin@astro.unistra.fr}\\
       Universit\'e de Strasbourg, CNRS, Observatoire Astronomique de Strasbourg, UMR 7550, F-67000 Strasbourg, France}

\date{Accepted 2018 June 8;
      Received 2018 June 7;
      in original form 2018 April 12}

\pagerange{\pageref{firstpage}--\pageref{lastpage}}
\pubyear{2017}

\begin{document}

\maketitle

\begin{abstract}
We create the first broadband polarization spectrum of an active galactic nucleus (AGN) by 
compiling the 0.1 -- 100~$\mu$m, 4.9~GHz and 15~GHz continuum polarization of NGC~1068 from more than 50 
years of observations. Despite the diversity of instruments and apertures, the observed spectrum of 
linear continuum polarization has distinctive wavelength-dependent signatures that can be related 
to the AGN and host galaxy physics. The impact of the Big Blue bump and infrared bump, together
with electron, Mie scattering, dichroism and synchrotron emission are naturally highlighted in polarization,
allowing us to reveal the type-1 AGN core inside this type-2 object with unprecedented precision. 
In order to isolate the AGN component, we reconstruct the spectral energy distribution of NGC~1068 and 
estimate the fraction of diluting light in the observed continuum flux. This allows us to clearly
and independently show that, in the case of NGC~1068, Thomson scattering is the dominant mechanism 
for the polarization in the optical band. We also investigate the effect of aperture on the 
observed polarization and confirm previous findings on the extension of the narrow line 
region of NGC~1068 and on the B-band and K-band polarization from the host. Finally, we do not 
detect statistically significant aperture-corrected polarimetric variations over the last 52 years, 
suggesting that the parsec-scale morphological and magnetic geometries probably remained stable for
more than half a century.
\end{abstract}


\label{firstpage}

\section{Introduction}
\label{Introduction}
NGC~1068, also commonly known as M77, is one of the most studied active galactic nuclei (AGN). Its proximity to Earth 
($z$ = 0.003793, $d$\footnote{Redshift-independent distance computed by the NASA/IPAC Extragalactic Database 
from 11 distances in the literature.} = 10.582$^{+5.418}_{-3.402}$~Mpc) and its bolometric luminosity ($\sim$2$\times$10$^{45}$erg~s$^{-1}$, 
\citealt{Honig2008,Raban2009}) make NGC~1068 an ideal target for spectroscopic, polarimetric and high resolution 
angular imaging studies. It allows high signal-to-noise ratio observations \citep[e.g.,][]{Low1971,Wilson1997,Alexander1999,Bauer2015},
which is imperative to distinguish the central AGN from its host spiral galaxy \citep{Balick1985}. It is an essential 
point as the properties of the intermediate region between the inner AGN dust material and the outer circumnuclear 
starforming regions of NGC~1068 can probe in great details the co-evolution of the AGN and its host galaxy 
\citep{Vollmer2008,Schartmann2009,Schartmann2010}.

The AGN in NGC~1068 was the key-point of the Unified Model for radio-quiet objects\footnote{We acknowledge
that the spectropolarimetric study of the radio galaxy 3C~234 preceded the work on NGC~1068, revealing a type-1 spectrum in polarized 
light together with a high polarization degree oriented perpendicular to the radio axis \citep{Antonucci1982,Antonucci1984}.
However 3C~234 is a radio-loud AGN, while NGC~1068 is radio-quiet.} such as postulated by \citet{Antonucci1993}. This zeroth 
order geometrical scheme has proven to be particularly successful in explaining a large fraction of observational 
features in many radio-quiet AGN thanks to a critical point: the nuclear orientation of the central engine. Depending
on the inclination of the observer with respect to the polar axis of the object, defined as the direction towards which 
the outflows of the AGN are directed, a variety of emission lines can be detected. \citet{Khachikian1974} defined two 
types of Seyferts galaxies. In their nomenclature Seyfert-2s show narrow permitted and forbidden emission lines while 
Seyfert-1s show broad permitted emission lines in addition to the Seyfert-2s lines. The reason for the disappearance 
of the broad permitted emission lines in types-2s was a crucial question that was solved largely thanks to observations 
of NGC~1068. \citet{Antonucci1985} have shown that high-resolution (5 -- 10\AA), high signal-to-noise ratio polarization 
spectra of the nucleus of NGC 1068 are necessary to uncover broad Balmer lines and Fe~{\sc ii} emission. Those lines have 
been found to have an intrinsic polarization $>$15\% at approximately the same position angle as that of the continuum, and 
their line wings appear broadened in the polarized flux spectra only. It was a definite set of proofs that a Seyfert-1 nucleus
is hiding inside NGC~1068. The polarization angle, perpendicular to the axis of radio emission, is the strongest evidence 
for polar scattering of inner AGN photons, imprinting the polarized spectrum with the broad emission line. To explain 
the absence of broad lines in total flux and the orientation of the polarization angle, the most convenient way is to
postulate the existence of a reservoir of circumnuclear dust that is enshrouding the nucleus along the equatorial 
plane. According to the inclination of the observer the central engine can be directly observed (Seyfert-1s) or it 
is obscured by dust (Seyfert-2s). It naturally explains the disappearance of the broad permitted emission line that 
can only be detected in polarized flux. The theory was also confirmed in radio-loud AGN \citep{Antonucci1984,Barthel1989}, 
leading to a unification of radio-loud AGN which includes radio galaxies, quasars, and blazars \citep{Urry1995}.

Polarimetric observations of NGC~1068 had a significant impact on our global comprehension of AGN but many questions 
that can be solved with polarimetry remain open. The systematic search for hidden type-1 cores in Seyfert-2s allows
one to test whether all AGN are fundamentally the same in terms of physical components. Optical spectropolarimetry 
revealed broad components of Balmer lines in many type-2s but a fraction of them are reluctant to show broad 
permitted emission lines in their linearly polarized spectrum \citep[see, e.g.,][]{Young1995,Young2000,Moran2000,Tran2003,Ramos2016}. 
New high resolution, high signal-to-noise ratio polarimetric observations of AGN on modern telescopes with large 
mirrors are necessary to test if some type-2 AGN genuinely lack a type-1 core. In addition, high angular resolution, 
high contrast polarimetric imaging using adaptive optics have shown that nearby bright AGN can be observed at 
phenomenal resolutions in the near-infrared \citep[e.g.,][]{Gratadour2015,Lopez2015}. \citet{Gratadour2015} used the 
extreme adaptive optics system on the SPHERE instrument at the Very Large Telescope to observe NGC~1068 in the
H (1.65$\mu$m) and K' (2.2$\mu$m) bands, and achieved polarimetric images with resolution 0.068$"$ ($\sim$ 4~pc at 
$d \approx$ 14~Mpc, which is the distance usually chosen for this object). By doing so, the authors revealed a compact 
elongated\footnote{Atacama Large Millimeter/submillimeter Array (ALMA) observations have revealed that the torus has 
an extension of $\sim$5x10pc at 432~{$\mu$}m \citep{Garcia2016,Gallimore2016,Imanishi2018}. The structure suggested 
by \citet{Gratadour2015} should be taken as an upper limit due to the methodology used by the authors.} (20x60~pc) 
structure, tracing the scattering regions at the center of NGC~1068, together with the already seen hourglass-shaped 
polar winds \citep[see, e.g.,][]{Capetti1995a,Capetti1995b,Kishimoto1999}. The later author pushed the polarimetric 
study of the ultraviolet linear polarization of the polar outflows up to the determination of the location of the 
nucleus of NGC~1068 and figured out the three-dimensional structure of its winds.

There is a lot more to be discovered by polarimetry. One can test the geometry of the region responsible for the emission 
of the broad permitted line \citep{Smith2002}. It should be possible to determine the spin, mass and inclination of the central 
supermassive black hole thanks to X-ray polarimetry \citep{Dovciak2004,Schnittman2009,Schnittman2010,Dovciak2011}. The 
composition and location of almost all the AGN components can be probed independently of the fact that they are obscured 
by dust or gas. Even the presence of aborted jets may be detected. To reach this goal, broadband polarimetry is needed.
\citet{Ramos2016} have clearly showed that information can be missed if we focus on a too narrow waveband. But is there 
enough published data to test the Unified Model of AGN at all wavebands? Do we need further instruments on large class 
telescopes? What is the broadband continuum polarization spectrum of radio-quiet AGN and what are the wavebands to be 
still explored? 

To answer those questions, we present a thorough compilation of all the continuum polarimetric observations of NGC~1068 
in order to create, for the first time, a combined spectrum of the linear polarization of an archetypal type-2 AGN. 
We use the fact that no other AGN have as much published polarimetric data as NGC~1068 to estimate the impact of several
observational constraints on the detected polarization levels. We also investigate the time-evolution of polarization 
over $\sim$52 years and check whether NGC~1068 has evolved in terms of geometry. By doing so we aim at making a strong 
case for future polarimetric instruments and programs. In this paper, we compile the published polarimetric data on 
NGC~1068 in Sect.~\ref{data:catalog} and build the combined polarized spectrum of NGC~1068 in Sect.~\ref{data:POL_spectrum}. 
We investigate the effect of apertures in Sect.~\ref{data:Aperture} before computing the global spectral energy 
distribution (SED) of the system in Sect.~\ref{data:SED}. We use this multi-component SED to correct the continuum polarization
from starlight dilution in Sect.~\ref{data:Correction} and demonstrate the predominance of electron scattering. Finally, 
we examine the temporal evolution of the linear polarization of NGC~1068 in Section.~\ref{data:Time_evolution}. We discuss
our results and the important observations to be made in the future in Sect.~\ref{Discussion} before concluding our paper 
in Sect.~\ref{Conclusions}.

\section{The data}
\label{data}
To gather the polarimetric observations published in refereed papers, the SAO/NASA Astrophysics Data System (ADS)
was extensively used. The SAO/NASA ADS is a Digital Library portal for researchers in Astronomy and Physics, operated 
by the Smithsonian Astrophysical Observatory (SAO) under a NASA grant; its website is accessible here: 
\url{http://adsabs.harvard.edu/}. Keywords and filtering by object name (NGC~1068 and its alternative names) allowed
for easier detection of relevant papers. Discussions with experts in the field, acknowledged at the end of this paper,  
also helped to find obscure publications. Several journal were also contacted to retrieve non-indexed results mentioned
in several publications (Astronomical Journal, Astronomical Circular).

\subsection{The catalog}
\label{data:catalog}

\begin{table*}
  \centering
  \begin{tabularx}{\textwidth}{p{0.18\linewidth}p{0.3\linewidth}p{0.2\linewidth}p{0.15\linewidth}p{0.02\linewidth}}
    \hline
    \textbf{Reference} & \textbf{Instrument} & \textbf{Waveband} & \textbf{Aperture} & \textbf{Imaging} \\
    \hline
    \citet{Dibai1966} & integrating electropolarimeter of the Crimean Astrophysical Observatory & U, B, V filters & 15$"$, 25$"$ & ~ \\
    \citet{Dombrovskii1968} & electropolarimeter of Leningrad University & U, B, V, R system, ZhS-18 and SZS-22 filters & 13$"$, 26$"$ & ~ \\
    \citet{Kruszewski1968} & Catalina 154-cm Telescope & U(2.78$\mu$m$^{-1}$), H(2.33), G(1.93), O(1.56), R(1.21), and I(1.06) & 9.2$"$, 10.2$"$ & ~ \\
    \citet{Visvanathan1968} & 60- and 100-inch telescopes at Mount Wilson & U, B, V, R system & 12.5$"$, 19.0$"$, 28.0$"$ & ~ \\
    \citet{Kruszewski1971} & Catalina 154-cm Telescope & U(2.78$\mu$m$^{-1}$), H(2.33), G(1.93), O(1.56), R(1.21), and I(1.06) & 6.8$"$, 9.2$"$, 10.2$"$, 15$"$, 30.6$"$ & ~ \\
    \citet{Knacke1974} & 1.3-m telescope of the Kitt Peak National Observatory & 3.5$\mu$m ($\Delta\lambda$=0.6$\mu$m), 10.2$\mu$m ($\Delta\lambda$=6$\mu$m), and 18.4$\mu$m ($\Delta\lambda$=1$\mu$m) & 12$"$ & ~ \\ 
    \citet{Angel1976} & 90 inch (2.3 m) telescope of Steward Observatory & 3200 -- 8600\AA & 2$"$ & ~ \\    
    \citet{Dyck1976} & \textit{unknown} & 1.2$\mu$m, 2.2$\mu$m & 5.4$"$, 11$"$ & ~ \\
    \citet{Elvius1978} & 72-inch Perkins Telescope of the Ohio Wesleyan Observatory & Schott UG1 filter & 13$"$-long slit & ~ \\  
    \citet{Lebofsky1978} & Steward Observatory 90-inch telescope & J, H, K, L' filters & 4$"$, 6$"$, 8$"$ & ~ \\    
    \citet{Wilson1982} & Very Large Array & 15~GHz & 0.7$"$ & \ding{51} \\    
    \citet{Martin1983} & Steward Observatory 2.3 m, Las Campanas 1.0 and 2.5 m, Kitt Peak 1.3 and 2.1 m, and University of Western Ontario 1.2 m telescopes & C500 filter (blue -- green) & 5$"$ & ~ \\  
    \citet{McLean1983} & 3.9-m Anglo-Australian Telescope & 3900 -- 8000\AA & 1.7$"$x2.2$"$, 2$"$x2.5$"$ & ~ \\     
    \citet{Miller1983} & Lick 3-m Shane Telescope & 3500 -- 5300\AA & 2.8$"$ & ~ \\
    \citet{Wilson1983} & Very Large Array & 4.9~GHz & 1$"$ & \ding{51} \\    
    \citet{Aitken1984} & 3.9-m Anglo-Australian Telescope and 3.0-m Infrared Telescope in Hawaii & 8 -- 13.1$\mu$m & 4.2$"$, 5.6$"$, 15$"$, 50$"$ & ~ \\    
    \citet{Antonucci1985} & Lick 3-m Shane Telescope Image Dissector Scanner & 3500 -- 7000\AA & 2.8$"$ & ~ \\
    \citet{Bailey1988} & 3.9-m Anglo-Australian Telescope & 0.36 -- 4.8$\mu$m & 4.5$"$, 6.0$"$ & ~ \\
    \citet{Scarrott1991} & 3.9-m Anglo-Australian Telescope & V, K filters & 1.6$"$, 2.8$"$, 4.5$"$ & \ding{51} \\
    \citet{Code1993} & Wisconsin Ultraviolet Photo Polarimeter Experiment & 1500 -- 3200\AA & 6$"$x12$"$ & ~ \\ 
    \citet{Antonucci1994} & Hubble Space Telescope Faint Object Spectrograph & 1575 -- 3300\AA & 0.3$"$, 1$"$, 4.3x1.4$"$ & ~ \\ 
    \citet{Capetti1995a} & Hubble Space Telescope Faint Object Camera and Wide Field Planetary Camera & 2700 -- 3700\AA, 5000 -- 6000\AA & 2.8$"$ & \ding{51} \\
    \citet{Capetti1995b} & Hubble Space Telescope COSTAR-corrected Faint Object Camera & 2400 -- 2700\AA & 2.8$"$ & \ding{51} \\   
    \citet{Tran1995} & Lick 3-m Shane Telescope & 4560 -- 7355\AA, 4600 -- 7400\AA, 3315 -- 4400\AA & 2.4$"$-slit & ~ \\    
    \citet{Young1995} & 3.8 m United Kingdom Infrared Telescope and the CGS4 spectrometer & 0.46 -- 0.77$\mu$m, 1.18--1.38$\mu$m and 1.66--2.07$\mu$m & 3.08x3.0$"$ & ~ \\    
    \citet{Packham1997} & IR imaging polarimeter at the Anglo-Australian Telescope & J, H, K$_{\rm n}$ filters & 2.0$"$, 4.5$"$, 6.0$"$ & \ding{51} \\
    \citet{Alexander1999} & United Kingdom Infrared Telescope, CGS4 spectrometer and IRPOL2 & 1.05 -- 1.35$\mu$m & 1.23$"$x6.7$"$ & ~ \\ 
    \citet{Lumsden1999} & 3.9-m Anglo-Australian Telescope & J, H, K$_{\rm n}$, N filters & 2.0$"$, 4.5$"$, 6.0$"$ & \ding{51} \\
    \citet{Simpson2002} & NICMOS Camera 2 on the Hubble Space Telescope & 2$\mu$m & 3.0$"$ & \ding{51} \\
    \citet{Watanabe2003} & 3.8 m United Kingdom Infrared Telescope & 0.46 -- 0.90$\mu$m, 0.92--1.80$\mu$m and 1.88--2.50$\mu$m & 3.5$"$ & ~ \\
    \citet{Packham2007} & Gemini North 8.1-m telescope & 9.7$\mu$m & 1.7$"$x1.2$"$ & \ding{51} \\
    \hline
  \end{tabularx}
  \caption{Catalog of published polarimetric measurements of NGC~1068. The first 
	  column is the reference paper, the second column is the instrument 
	  used for the measurement, the third column is the waveband or filters used 
	  during the observation, the fourth column is the observation aperture
	  (in arcseconds), and the fifth column indicates if polarimetric images were
	  taken.}
  \label{Tab:Catalog}
\end{table*}

\begin{table*}
  \centering
  \begin{tabularx}{\textwidth}{p{0.18\linewidth}p{0.3\linewidth}p{0.2\linewidth}p{0.15\linewidth}p{0.02\linewidth}}
    \hline
    \textbf{Reference} & \textbf{Instrument} & \textbf{Waveband} & \textbf{Aperture} & \textbf{Imaging} \\
    \hline
    \citet{Mason2007} & IRPOL2 spectropolarimetry module and CGS4 on the 3.8m UK Infrared Telescope & 3.10 -- 3.67$\mu$m & 0.6$"$-slit & ~ \\
    \citet{Lopez2015} & MMT-Pol on the 6.5-m MMT & J', K' filters & 0.2$"$, 0.5$"$, 2.0$"$ & \ding{51} \\
    \citet{Lopez2016} & CanariCam on the 10.4-m Gran Telescopio CANARIAS & 8.7$\mu$m, 10.3$\mu$m, 11.3$\mu$m, 11.6$\mu$m & 0.4$"$, 2.0$"$ & \ding{51} \\ 
    Grosset et al. (in prep.) & SPHERE/VLT & H and K' bands & 0.2$"$, 0.5$"$, 1.0$"$, 2.0$"$, 3.0$"$, 4.0$"$, 5.0$"$ & \ding{51} \\  
    Lopez-Rodriguez et al. (in prep.) & HAWC+ on the 2.5-m SOFIA telescope & 53$\mu$m, 89$\mu$m & 5.0$"$, 8.0$"$ & \ding{51} \\   
    \hline
  \end{tabularx}
  \caption{Tab.~\ref{Tab:Catalog} continued.}
  \label{Tab:Catalog2}
\end{table*}

The final catalog of polarimetric\footnote{We focus on NGC~1068 linear polarization. Circular polarization, at least in the 
near-ultraviolet, optical and near-infrared bands, has been the subject of an unsolved debate. One paper claimed a non-detection 
\citep{Landstreet1972}, another a non-significant detection \citep{Gehrels1972} and two others a clear detection 
\citep{Nikulin1971,Angel1976}.} observations of NGC~1068 is presented in Tab.~\ref{Tab:Catalog}. It contains 34 publications spanning 
over more than 50 years (1965 -- 2017, observational dates, not paper publication time-stamp). All papers are accounting for instrumental
polarization, together with interstellar polarization. Due to the high Galactic latitude of NGC~1068 (-51.93$^\circ$), Galactic contamination 
by diffuse interstellar grains is not expected to impact the measured polarization in the ultraviolet -- far-infrared band 
\citep{Prunet1998}. However a large fraction of authors did not corrected their data for dilution/contamination of the polarization by the host
galaxy (exceptions include, but are not limited to, the work by \citealt{Miller1983}, \citealt{Antonucci1985}, \citealt{Kishimoto1999} and
\citealt{Lopez2015}). The dust lane and host galaxy have a visual extinction A$_{\rm V} \sim$ 9~mag to the core of NGC~1068, producing 
an aperture-dependent level of dilution to be corrected at ultraviolet, optical and infrared wavelengths. Unfortunatelly this correction 
was not achieved by all the authors. To have a consistent set of measurements, the polarized values extracted from the aforementioned papers 
are the one that were not corrected for starlight dilution (but see Sect.\ref{data:SED} for the correction of polarization by removing the
host component). The measurements were taken with a variety of instruments, listed in the second column of Tab.~\ref{Tab:Catalog}. Observations
were made in different wavebands, from the ultraviolet using the Wisconsin Ultraviolet Photo Polarimeter Experiment \citep{Code1993} to 
the far-infrared using the SOFIA High-resolution Airborne Wideband Camera-plus (HAWC+, Lopez-Rodriguez, E., private communication). 
Only two polarimetric observations have been published in the radio band, an upper limit at 15~GHz \citep{Wilson1982} and a debated 
measurement at 4.9~GHz \citep{Wilson1983}. The bulk of observations were taken in the optical and near-infrared bands since they remain the 
easiest bands for ground observations. A variety of slits and circular apertures were used, depending on the technology available at that time 
(fourth column of Tab.~\ref{Tab:Catalog}). Finally, we have identified observations that were achieved in imaging modes. For these cases the 
authors had polarization maps and could have, in principle, varied the aperture to remove the contribution of the host galaxy. 

\begin{figure}
    \includegraphics[trim = 0mm 0mm 0mm 0mm, clip, width=8.5cm]{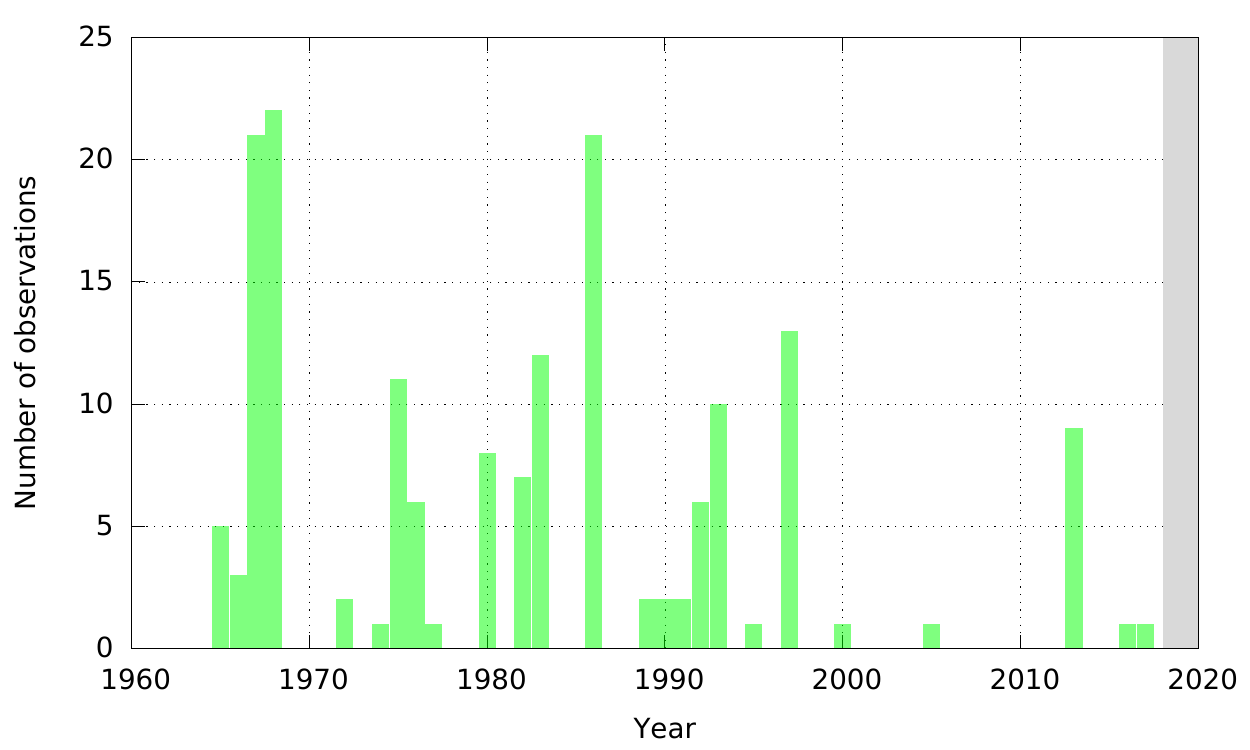}
    \caption{Number of observations per year dedicated to 
	    measure the broadband continuum polarization
	    of NGC~1068. The shaded area corresponds to the 
	    forthcoming years 2019 -- 2020.}
    \label{Fig:Histogramme}
\end{figure}

In Fig.~\ref{Fig:Histogramme}, we present the temporal distribution of the number of polarimetric observations listed in our catalog.
The very first polarimetric observation of NGC~1068 listed in our sample goes back to 1965 -- 1966 \citep{Dibai1966}. A year later,
\citet{Dombrovskii1968} quoted polarimetric observations ``with a somewhat smaller aperture than ours'' that might also have been achieved
by Merle F. Walker in 1964 but the references listed in \citet{Dombrovskii1968} point towards papers that are either about other galaxies 
(M33, \citealt{Walker1964}) or do not exist at all. A similar reference to the work by Walker (Astron.J., 1964, 69, 682) is mentioned
in \citet{Elvius1965} but the related paper could not be found. Hence, while we acknowledge that earlier polarimetric observations might 
have been achieved, we start our catalog in 1965 -- 1966 due to the lack of open-access publications. 

The interest of the community for the polarimetric signature of NGC~1068 and similar galactic nuclei grew fast, with more than 50 observations
achieved before 1970. The goal was to explore the origin of the observed polarization. Synchrotron emission was one of the two 
mechanisms (together with scattering) suggested to explain the high ultraviolet polarization in its nucleus \citep{Elvius1965}. A few 
additional polarimetric measurements of this AGN occurred until the advent of the 3.9-m Anglo-Australian Telescope and the Lick 3-m Shane 
Telescope, which gave a new kick to observations in 1983 thanks to their large mirrors and up-to-date polarimeters. The true scattered 
polarization of NGC~1068 was estimated to be much higher than previously thought thanks to the careful removal of starlight from the 
host galaxy by \citet{Miller1983}. In addition, it was found that the position angle of the optical continuum radiation is perpendicular
to the axis of radio emission (18$^\circ$ $\pm$ 5$^\circ$, \citealt{Wilson1982}). After the breakthrough achieved in 1985, where 
\citet{Antonucci1985} proved that a Seyfert-1 nucleus lies hidden in the core of NGC~1068, and that the origin of the continuum and 
broad-line polarization is due to scattering, the number of observations decreased. It was only in the mid of the 90's, when the 
Unified Model of AGN was synergized, confirmed and then reviewed \citep{Antonucci1993}, that the community acquired a few more 
polarimetric observations of NGC~1068. Since the new millennium, only a dozen of polarimetric observations of this AGN have been 
achieved on 10-m class telescopes. Another observational gap could appear between the era of 10-m and 30-m class telescopes, at 
least partly driven by the relative lack of polarimeters on such large telescopes.

\subsection{Broadband continuum polarization of NGC~1068}
\label{data:POL_spectrum}

\begin{figure*}
    \includegraphics[trim = 0mm 0mm 0mm 0mm, clip, width=17cm]{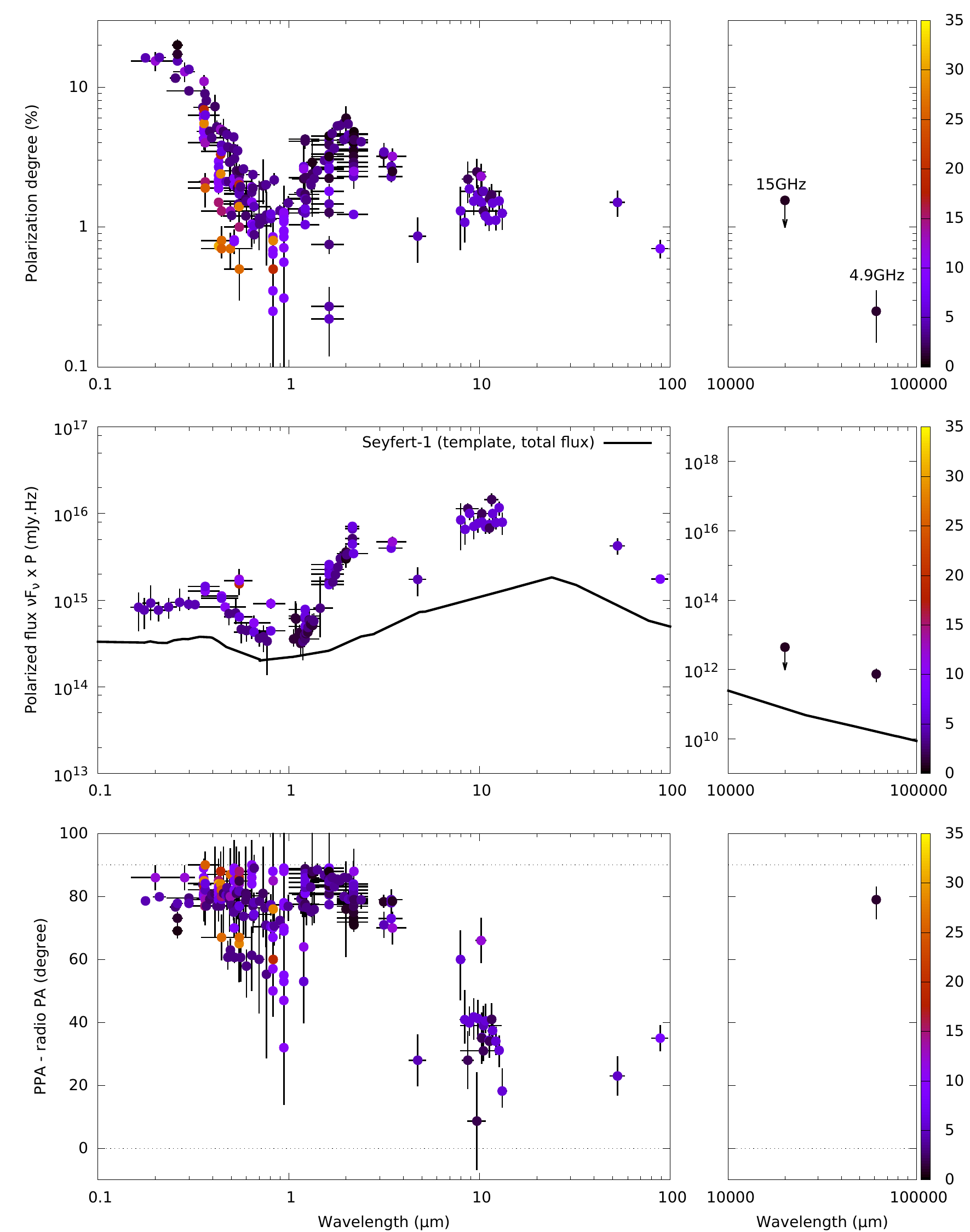}
    \caption{Broadband 0.1 -- 100~$\mu$m, 4.9~GHz and 15~GHz continuum polarization
	    of NGC~1068 measured from various instruments and 
	    apertures (color-coded, in arcseconds). Instrumental 
	    and interstellar polarization have been accounted 
	    for but contribution from the host galaxy is still
	    present. See text for details. Top: polarization degree,
	    middle polarized flux ($\nu$F$_\nu$, in mJy.Hz, times $P$),
	    bottom: polarization position angle (PPA) minus the 
	    parsec-scale radio position angle (PA). The 
	    Seyfert~1 total flux template presented in the polarized flux 
	    figure (middle) was shifted downward for better visibility.}
    \label{Fig:POL_spectrum}
\end{figure*}

We compiled all the linear continuum polarization data of NGC~1068 in Fig.~\ref{Fig:POL_spectrum}. When continuum polarization measurements 
were not estimated by the authors we used WebPlotDigitizer to synthesize the polarization spectrum and extract the relevant numbers. 
WebPlotDigitizer (\url{https://automeris.io/WebPlotDigitizer/}) is a polyvalent and free software developed to facilitate easy and accurate 
data extraction from spectra and it has already been used in \citet{Marin2017} to reconstruct the ultraviolet spectropolarimetric spectrum of
NGC~1068. 

Our final spectrum spans from almost 0.1~$\mu$m to 100~$\mu$m, together with two additional measurements at 4.9~GHz and 15~GHz, with 
a variety of apertures, signal-to-noise ratios and spectral resolutions. In this figure, we did not correct the polarization levels for the 
presence of diluting starlight emission originating from the host galaxy that may have a significant effect depending of the aperture of the 
observation/slit. This will be investigated in further details in Sect.~\ref{data:Aperture}. We plotted the continuum linear polarization in 
log scale to better contrast the fractional contribution of unpolarized light. The polarized flux is extracted from publications and is 
compared to a normal type-1 AGN total flux spectrum to investigate how the cross section of the scatterer is changing with wavelength. 
The type-1 template we use was compiled by \citet{Prieto2010} and is the averaged SED of the high spatial resolution SEDs of NGC~3783, 
NGC~1566 and NGC~7469. The template was re-scaled in order to be easily comparable to the polarized flux of NGC~1068. Finally, the 
polarization position angle has been subtracted from the parsec-scale radio position angle to check whether the polarization angle is parallel
or perpendicular to the axis of the radio source associated with the galaxy \citep{Antonucci1993}. We use the parsec-scale radio position 
angle estimated by \citet{Wilson1982} but we acknowledge the fact that the position angle is almost 0$^\circ$ at sub-arcesond scales 
\citep{Muxlow1996}. Our choice to use the parsec-scale value is coherent as the bulk of published polarimetric data having apertures
larger than 1$"$.

We can see from Fig.~\ref{Fig:POL_spectrum} that the compiled polarization spectrum of NGC~1068 shows a coherent energy-dependent behavior 
despite the multiple instruments, observational apertures and observational dates. The linear continuum polarization is the highest in the 
ultraviolet band where starlight emission is weak: the starlight fluxes of spiral galaxies are about three orders of magnitude lower at 
0.1~$\mu$m than at 1~$\mu$m \citep{Bolzonella2000,Siebenmorgen2007}. With increasing diluting fluxes from starlight, the continuum polarization
of NGC~1068 decreases from $\sim$15\% at 0.1 -- 0.2~$\mu$m to $\sim$1\% at 0.8 -- 0.9~$\mu$m. The polarized flux, despite being not as well 
sampled as the polarization degree due to the lack of reported flux measurements, clearly shows the turnover of host dominance. At ultraviolet
wavelengths the polarized flux is constant while it decreases sharply in the optical, dipping at $\sim$1$\mu$m. The dip in polarization at 
$\sim$1$\mu$m is due to the maximum of starlight contribution that almost cancel the observed polarization \citep{Bolzonella2000}. This 
polarization dip is also consistent with the transition waveband between the ``Big Blue Bump'' and the ``infrared bump'' detected in the 
spectral energy distribution of AGN. The Big Blue Bump is due to multi-color black body emission from the accretion disk and the infrared bump
is attributed to thermal emission from dust \citep{Sanders1989,Wilkes2004}. The inflection between the two bumps, related to dust sublimation 
at temperature 1500 -- 2000K, occurs at 1 -- 2$\mu$m, similarly to the onset of the infrared polarized peak. As it can be seen in 
Fig.~\ref{Fig:POL_spectrum}, both the polarization degree and the polarized flux are strongly increasing in the near-infrared band. 
Interestingly, it appears that the polarized flux spectrum rises much more sharply in the near-infrared than the average type-1 SED that 
we have plotted. In this case, the orientation difference between NGC~1068 and the averaged SED of \citet{Prieto2010} certainly plays a role 
as polarization by polar scattering transitions in the infrared to polarization due absorption/emission by aligned grains (this will be developed 
in the next paragraph). Polar scattering provides a similar face-on view to what we see in type-1 AGNs, but in the case of dichroic polarization, 
we are seeing the transmitted (or emitted) light from the edge on torus dust distribution.

At longer wavelengths the host emission decreases and the infrared polarization shows a second maxima at wavelengths 2 -- 3~$\mu$m. The absence of 
variation seen in the polarization position angle of light (that remains perpendicular to the axis of radio emission from the ultraviolet band to $\sim$ 
3~$\mu$m) indicates that most of the polarization is due Thomson and Mie scattering in the polar region. However, at 4 -- 5~$\mu$m, the polarization
position angle switches from perpendicular to almost parallel. This effect was already highlighted and discussed by \citet{Bailey1988}. The authors
have shown that the angle rotation is not a consequence of the observed forbidden lines flux included in the passbands of the filters used, which
would have the effect of pulling the position angle towards the larger forbidden line value, but it is in fact a real feature of the continuum 
polarization. While the polarization in the ultraviolet, optical and near-infrared bands is caused by polar scattering by electron and dust 
grains, the polarization at $\lambda \ge$ 4~$\mu$m is most likely due to thermal emission from 100~pc dust grains aligned by large-scale magnetic
fields \citep{Efstathiou1997,Lopez2015}. Since most the dust is concentrated along the AGN equatorial plane, the polarization angle aligns with 
the radio axis and becomes parallel. The sharp decrease of polarized flux we observe at 4 -- 5~$\mu$m confirms the variation of the cross section 
of the scatterer. We are then able to trace the exact wavelength at which polar scattering becomes inefficient in NGC~1068. 

In mid- and far-infrared bands, 10~$\mu$m polarization observations at subarcsecond resolution showed that the core of NGC~1068 becomes 
consistent with zero measured polarization; the main contributor to polarization is the extended emission that comes from the polar material / GMC 
(giant molecular clouds) interactions in the Northern ionization cones \citep{Lopez2016}. Such discovery highlights the importance of correlating 
near- and mid-infrared data to draw conclusions. The polarized flux follows the dust emission by aligned grains and strongly increases between 10 
and 20~$\mu$m. The polarized flux appears to reach a maximum in the uncharted 20 -- 40~$\mu$m and far-infrared polarimetric measurements indicate 
that the polarized flux decreases in the 50 -- 100~$\mu$m waveband. 

Finally, in the radio band, two Very Large Array (VLA) measurements of the core polarization of NGC~1068 have been published. An upper limit was estimated
at 15~GHz and the 4.9~GHz apparent polarization of the central AGN component is at the limit of accuracy of the measurement technique. Nevertheless,
the polarized flux follows the expected trend of the type-1 total flux template with great consistency. The polarization degree is low, of the order 
of 0.25\% at 4.9~GHz and below 1.5\% at 15~GHz. Such low polarization degrees may be attributed to dilution by large quantities of thermal electrons 
as evidenced by the intense optical emission lines \citep{Kraemer1998}. The polarization angle could only be measured at 4.9~GHz and is essentially 
perpendicular to the source axis, similarly to the optical continuum polarization \citep[e.g.,][]{Martin1983}. We thus detect a second rotation of
the polarization position angle from the far-infrared to the radio domain. This rotation points toward a mechanism that is parallel to the polar 
magnetic fields. Accounting for the low polarization degree due to dilution by thermal electrons and the polarization position angle that is 
perpendicular to the radio-axis, electron-scattered synchrotron emission appears to be the most plausible scenario \citep{Gallimore2004,Krips2006}. 
This is in agreement with the work of \citet{Krips2006} who investigated several emission mechanisms to explain the millimeter-to-radio continuum
emission in NGC~1068. Indeed, the authors concluded that "the core fluxes indicate a turnover of the inverted cm- into a steep mm-spectrum at 
roughly 50~GHz which is most likely caused by electron-scattered synchrotron emission".

In conclusion, it is truly remarkable to observe that the whole polarized SED of NGC~1068 is following the averaged total flux type-1 SED extracted 
from \citet{Prieto2010}, proving that a type-1 core genuinely resides inside NGC~1068. We also note that the different emission and reprocessing 
mechanisms have a deep impact onto the polarization position angle, allowing us to probe the physics inside the core of obscured AGN.

\subsection{Impact of aperture onto the observed polarization}
\label{data:Aperture}

\begin{figure}
    \includegraphics[trim = 0mm 0mm 0mm 0mm, clip, width=8.5cm]{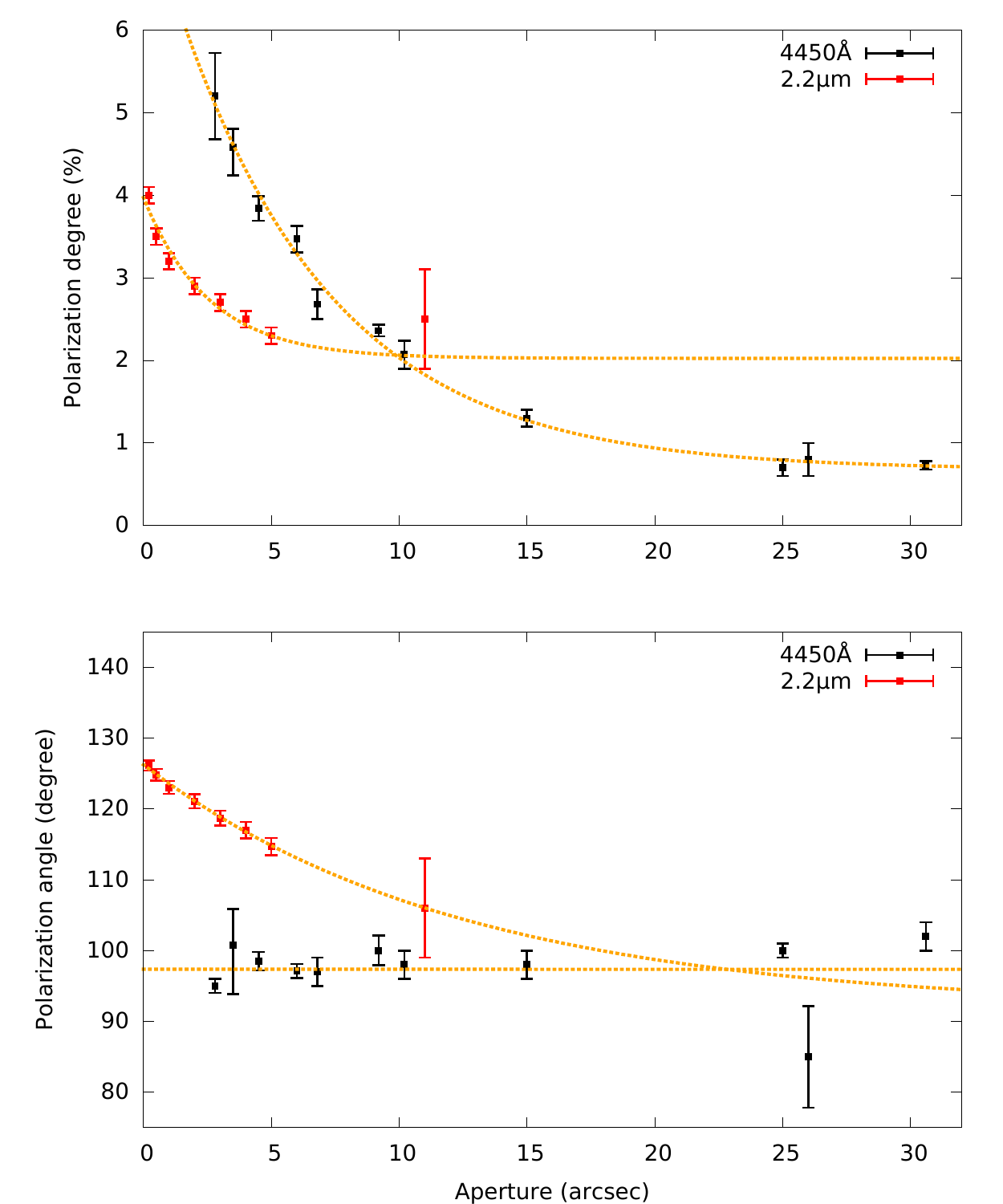}  
    \caption{Aperture effect on the measured continuum polarization
	    of NGC~1068 around 4450\AA~(in black) and around
	    2.2$\mu$m (in red). The first panel shows the variations
	    of the polarization degree, the second panel is for 
	    the polarization position angle. Fits to the data 
	    are shown in orange dashed lines.}
    \label{Fig:Aperture}
\end{figure}

\begin{table}
  \centering
  \begin{tabular}{l|r}
     P$_{\rm 4450}$ = 0.68+6.99*$\exp$(-0.16*Ap) & R$^2>$0.99 \\
     P$_{\rm 2.2{\mu}m}$ = 2.02+1.95*$\exp$(-0.39*Ap) & R$^2>$0.97 \\
     ~\\
     \hline
     ~\\
     $\Psi_{\rm 4450\AA}$ = -0.49.10$^{-4}$*Ap+97.37 & R$^2<$0.01 \\     
     $\Psi_{\rm 2.2{\mu}m}$ = 91.90+34.34*$\exp$(-0.08*Ap) & R$^2>$0.99
  \end{tabular}
  \caption{Fits to  variation of polarization degree $P$
	   and polarization position angle $\Psi$ as a function 
	   of aperture $Ap$ from Fig.~\ref{Fig:Aperture}.
	   The coefficient of determination R$^2$ is
	   indicated in the second column.} 
  \label{Tab:Fit}
\end{table}

It is well known that the measured AGN polarization depends on the aperture used \citep[see, e.g.,][]{Bailey1988}. The better we isolate 
the compact nucleus from the host starlight and starburst activity, the higher the ultraviolet, optical and near-infrared polarization 
degree\footnote{Thanks to high angular-resolution polarimetric observations, \citet{Packham2007} and \citet{Lopez2016} have shown that the 
10~$\mu$m polarization decreases with decreasing aperture. This is due to the complex combination of a) the extended emission from the 
wind-GMC interaction in the Northern ionization cones, b) the absorption polarization in the Southern cone, and c) the unpolarized core 
from self-absorbed dichroic emission from the torus.}. This effect is clearly visible in Fig.~\ref{Fig:POL_spectrum}, where the color-code 
highlights the fact that polarimetric measurements with large apertures are always smaller than polarimetric observations at the same wavelength
with a smaller observational aperture. Yet, since our catalog compiles all polarimetric data recorded for NGC~1068, we can investigate the impact
of aperture onto the resulting polarization with better accuracy. We isolate two regions from Fig.~\ref{Fig:POL_spectrum} where the polarization
degree is found to vary over several percentage points at a given wavelength: around 4450\AA~ (B-band) and around 2.2$\mu$m (K-band). The 
polarization dependency on the aperture is shown in Fig.~\ref{Fig:Aperture} (top: polarization degree, bottom: polarization angle). The aperture 
varies from 2.8$"$ to 30.6$"$ in the blue band and from 0.2$"$ to 11$"$ in the infrared, which corresponds to 205 -- 2250~pc and 14 -- 810~pc,
respectively. We see that the observed polarization degree indeed increases gradually as we get closer to the active nucleus. Dilution by the host
is relatively important for large apertures and the polarization degree exponentially increases when the observer reaches apertures lower than 
15$"$. We fitted the data point using exponential function that are summarized in Tab.~\ref{Tab:Fit}. All fits have a coefficient of determination
R$^2$, that is the proportion of the variance in the dependent variable that is predictable from the independent variable, superior to 0.97. 
However, there is no evidence for a variation in the polarization position angle at 4450\AA~as aperture increases. Overall, we find a
clear exponential dependence of the polarization properties with respect to the observational aperture. Our fits can then help to predict the 
expected polarization degree and angle at a given aperture for a given waveband. Additional polarimetric observations are needed in a large 
variety of wavebands to generalize our equations. 

The observed polarization plateaus and the rotation of the polarization position angle indicate the presence of large scale polarization that 
can only be attributed to the host itself. Scattering of starlight by dust, molecules and electrons in the galactic medium is known to produce a 
low amount of optical polarization that depends on the orientation of the host plane \citep{Scarrott1991,Simmons2000}. For an edge-on galaxy, the 
expected large scale polarization lies between 0.8 and 1.8\% \citep{Simmons2000}. The former value well corresponds to the plateau reached by 
the optical polarization curve in Fig.~\ref{Fig:POL_spectrum} (top). This polarization degree is in agreement with the optical linear 
polarization maps of NGC~1068 obtained by \citet{Scarrott1991}, who have shown that the polarization data from $r <$ 10$"$ traces the 
large-scale structure of the host. The spiral structure of the host, forming a roughly circularly symmetric pattern, almost certainly accounts 
for some of the decrease in polarization as the aperture size increases. At longer wavelength, near-infrared polarimetry of a normal spiral galaxy 
allowed \citet{Clemens2013} to show that the polarization fraction of late-type galaxies can reach up to 3\%, which also corresponds to the plateau 
of 2.2$\mu$m polarization we found. In conclusion, the diversity of apertures used to achieve polarimetric observations of NGC~1068 allowed us 
to confirm the amount of optical and near-infrared polarization from its host. 

Additionally, according to \citet{Schmitt1996} and \citet{Murayama1997} the optical size of the extended narrow line region in NGC~1068 is about 
900~pc (12.25$"$), which roughly corresponds to the inflexion point of the optical polarization fit as a function of aperture in Fig.~\ref{Fig:Aperture}. 
Since most of the low-aperture optical continuum polarization we observe from NGC~1068 is due to scattering of disk photons inside 
the polar outflows, and since we have demonstrated that large aperture polarization is dominated by the host, we can safely confirm that the extension
of the narrow line region probably stops before 1030~pc. This value is at a safe distance from the abrupt fall in the surface brightness profile
of NGC~1068 observed by \citet{Sanchez2004} and that is situated around 2.2~kpc from the center of the galaxy.

Consequently, despite the fact that the compiled continuum polarization spectrum of NGC~1068 presented in Fig.~\ref{Fig:POL_spectrum} suffers from the 
various apertures used, we know from Tab.~\ref{Tab:Catalog} that only a few observations were done with apertures larger than the estimated extension of 
the scattering outflows. This indicates that most of the data have indeed measured the polarization originating mainly from the first hundreds of parsecs 
surrounding the AGN core and that the features we see in Fig.~\ref{Fig:POL_spectrum} are not artifacts. Even if all the different apertures do play a role, 
they only contribute to a lesser extent to the characteristic wavelength-dependent polarization profile.

\subsection{Reconstructing the spectral energy distribution}
\label{data:SED}

\begin{figure}
    \includegraphics[trim = 0mm 0mm 0mm 0mm, clip, width=8.5cm]{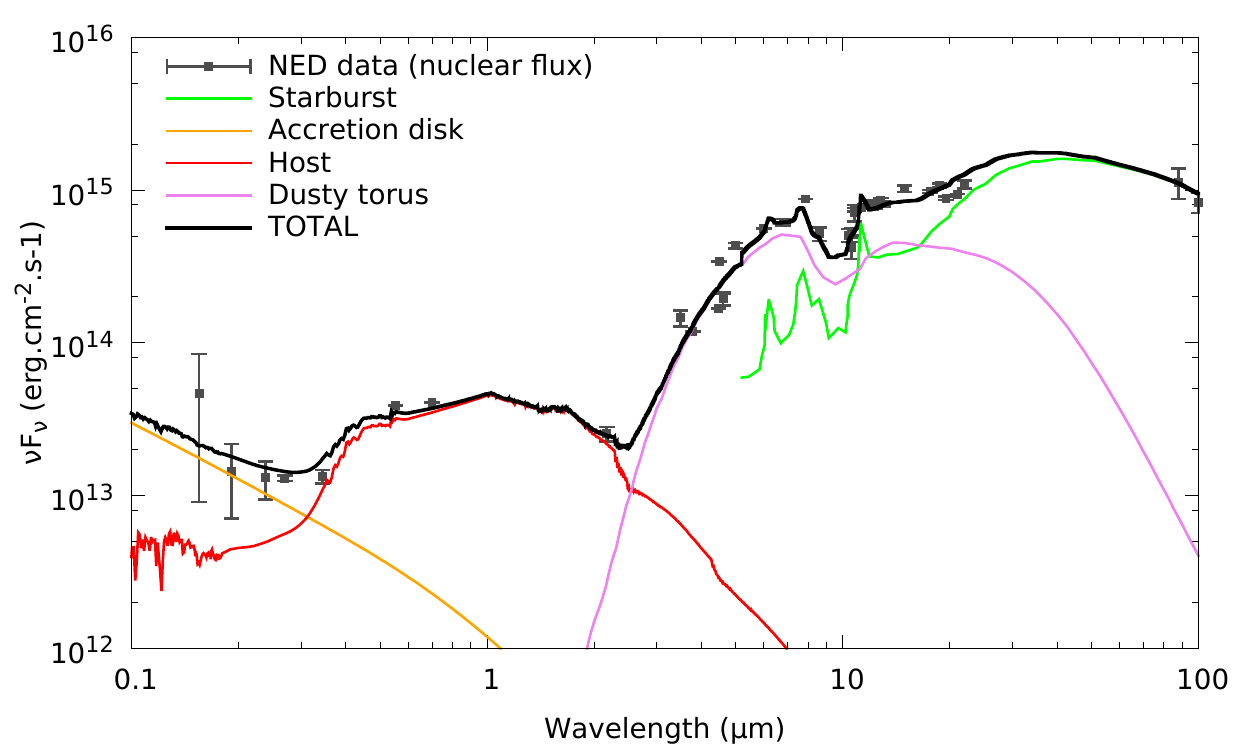}
    \caption{Observed nuclear fluxes of NGC~1068 (gray squares) extracted from the NASA/IPAC 
	     Extragalactic Database (NED). See text for details about the emissive components 
	     used to reproduce the broadband SED.}
    \label{Fig:SED}
\end{figure}

Correcting the continuum polarization from diluting light is a major complication. While it is important to estimate the instrumental polarization
and the contamination by interstellar polarization, it is crucial to remove the dilution by starlight in order to study the wavelength
dependence of polarization \citep{Miller1983}. To do so, the aforementioned authors carried out observations of M32, a morphologically classified
compact elliptical galaxy, in order to remove any unpolarized starlight that is dominating the continuum polarization of NGC~1068 longwards of 
4000\AA. By minimizing the stellar absorption features in the residuals of the M32/NGC~1068 flux ratio spectra, \citet{Miller1983} derived the 
following starlight fractions in the observed AGN continuum flux: 0.42 at 3600\AA, 0.63 at 4200\AA, 0.78 at 4600\AA, and 0.82 at 5075\AA ~(with 
5\% uncertainty). They corrected their linear polarization spectrum and found a wavelength-independent polarization of 16 $\pm$ 2\% in the 
3500 -- 5200\AA~waveband. However, to achieve so, high-resolution ($<$10\AA) spectropolarimetry is necessary. Most of the published polarimetric 
measurements of NGC~1068 were achieved in narrow-band filters, which prevent us to use this method. 

We thus opted for a similar procedure that rather focuses on the continuum flux of NGC~1068. We extracted from the NASA/IPAC Extragalactic Database 
all the nuclear observations of NGC~1068 and plotted them in Fig.~\ref{Fig:SED}. We focused on nuclear fluxes in order to better 
estimate the fraction of starlight in the observed continuum flux of NGC~1068. We reconstructed the 0.1 -- 100~$\mu$m spectral energy distribution
(SED) of NGC~1068 using usual AGN components: the scattered light of a thermally-emitting multi-color accretion disk, a component reproducing the 
infrared reemission of the circumnuclear AGN torus, a template for the host galaxy, and a template for the starburst light contribution. 

\begin{itemize}
\item To construct the accretion disk SED, we used the standard thin disk model \citep{Shakura1973}. The accretion disk spans from 1 to 1000 
gravitational radii and surrounds a 8.10$^6$ solar mass black hole \citep{Lodato2003} accreting at 0.4 times the Eddington rate \citep{Kumar1999}. 
The inclination of the disk was fixed to 85$^\circ$ according to the numerical reconstruction of NGC~1068 achieved by \citet{Fischer2013,Fischer2014}. 
The question about the true inclination of the system remains opened \citep{Marin2016} but we checked that varying those parameters only marginally 
influence the power-law shape of the scattered light of the disk SED observed in the ultraviolet-optical band. The variation are within the expected 
range of values (power-law index $\sim$ 1/3 due to the superposition of black-bodies). 
\item Dust reemission by the obscuring equatorial torus is simulated using the model presented by \citet{Fritz2006} to fit the infrared SED of NGC~1068.
The toroidal model has an aperture angle of 160$^\circ$, an optical depth of 8 at 9.7~$\mu$m, and is characterized by multiple grain temperatures set by 
thermal equilibrium equations. The torus extends up to 16.4~pc, has an outer-to-inner radii ratio of 20 and its dust density distribution varies both in 
the radial and in the altitude coordinates. We note that \citet{Fritz2006} used a homogeneous distribution of dust to describe the torus; in reality ALMA 
has shown that the outer edge of the torus is likely to be clumpy \citep{Garcia2016,Gallimore2016}. Its outer radius is of the order of 10pc but most of 
the obscuring material of the torus is concentrated around 5pc, according to the clumpy torus models \citep{Garcia2016}. This results in a model that may 
overestimate the far-infrared/millimeter emission that is, in any case, extended. 
\item A template for the host galaxy has been extracted from \citet{Bruzual1993} and was extended towards the ultraviolet and infrared bands by 
\citet{Bolzonella2000}. It corresponds to an archetypal Sbc barred spiral galaxy, which is consistent with the classification of the host of NGC~1068 
\citep{Balick1985}. The initial mass function by \citet{Miller1979} was used with an upper mass limit for star formation of 125~M$_\odot$. 
The database used to compute this SED includes only solar metallicity SEDs but it appears to have a minimum impact onto the resulting template \citep{Bolzonella2000}.
\item Finally, starburst activities are present in the central kilo-parsec of NGC~1068 \citep{Lester1987,Romeo2016}, imprinting the 5 -- 100~$\mu$m 
infrared band with strong features \citep{Thronson1989,Floch2001}. Following \citet{Fritz2006}, we included a starburst component needed to reproduce
the intensity of the polycyclic aromatic hydrocarbon features between 6 and 15~$\mu$m, as well as the depth of the silicate feature at 9.7~$\mu$m, and
the width, intensity and peak wavelength of the infrared bump. To do so, we include the contribution of the infrared spectrum of the starburst 
galaxy NGC~7714 that correctly reproduces the colder component of dust emission \citep{Fritz2006}.
\end{itemize}

Our final SED is presented in black in Fig.~\ref{Fig:SED} and it satisfactorily reproduces observations (in gray). This SED has been chosen to minimize the
differences between the model and the observed fluxes but we remark slight degeneracies due to the data errors bars, particularly in the ultraviolet and 
blue bands. Nevertheless, the shape of the near and mid-infrared spectrum is very well reproduced using a generic template for the host rather than using 
a specific galaxy observation such as M32. The transition between disk emission and torus emission happens at the expected wavelength and the host galaxy 
indeeds dominates the continuum flux of NGC~1068 longwards of 3000\AA.

\subsection{Polarization correction}
\label{data:Correction}

\begin{figure}
    \includegraphics[trim = 0mm 0mm 0mm 0mm, clip, width=8.5cm]{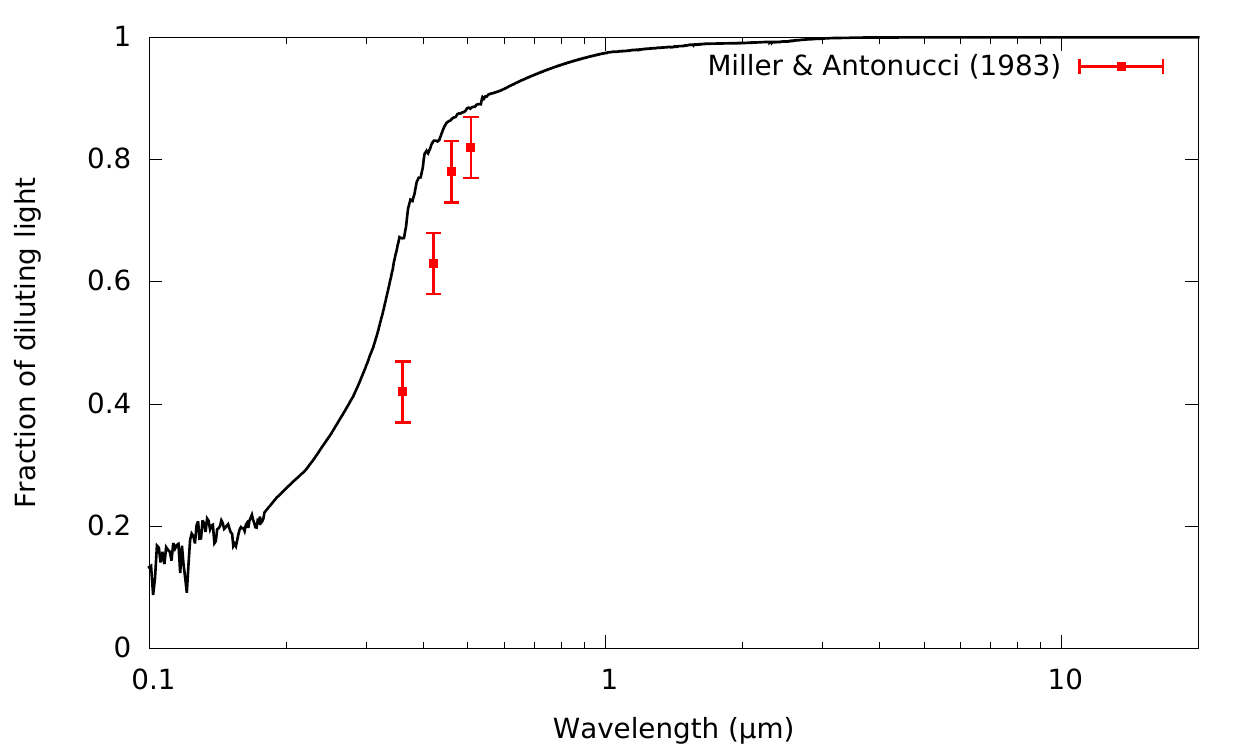}
    \caption{Derived fraction of diluting light (starlight plus dust emission 
	     components) in the observed continuum flux of NGC~1068. The previous 
	     estimation made by \citet{Miller1983} using a circular aperture of 
	     2.8$"$ is shown in red.}
    \label{Fig:Fraction}
\end{figure}

The overall agreement between the observed data points and our SED model allows us to derive the fraction of starlight plus dust 
emission components (dust-reprocessed accretion disk emission and starburst emission) in the observed continuum flux of NGC~1068. The 
flux ratio is shown in Fig.~\ref{Fig:Fraction}, together with the ratios derived by \citep{Miller1983} using the M32 template. Although 
the ratio are not the same (this being due to the two different approaches: spectropolarimetric fitting versus SED fitting, the former being
unfeasible in this paper), the shape of the wavelength-dependent ratios are distinctively similar. In our model, the host galaxy
starlight starts at bluer wavelengths, hence the shift by almost 1000\AA. We tried to reproduce the exact same flux ratios estimated by \citep{Miller1983}
by changing the normalization of our SED components but the observed data points were no longer matched in the optical band. We stress that this is
a logical outcome: first there is a difference in the apertures being used; second the two host galaxy SED used in \citet{Miller1983} and in our 
paper are not the same. In particular the choice of M32 may have had consequences on the spectropolarimetric analysis. M32 is an elliptical 
satellite galaxy of the M31 subgroup, together with NGC~205. A giant stream in the outer halo of M31, pointed in the direction of M32, was observed
by \citet{Ibata2001}, leading to potential tidal interaction between the galaxies \citep{Choi2002}. But, despite the morphological and chemical
differences between the two templates, the most important investigation is to test whether our SED and the fraction of starlight plus dust 
emission components in the observed continuum flux of NGC~1068 can reproduce the expected polarization thresholds quoted by \citet{Miller1983} 
and \citet{Antonucci1985}.

\begin{figure}
    \includegraphics[trim = 0mm 0mm 0mm 0mm, clip, width=8.5cm]{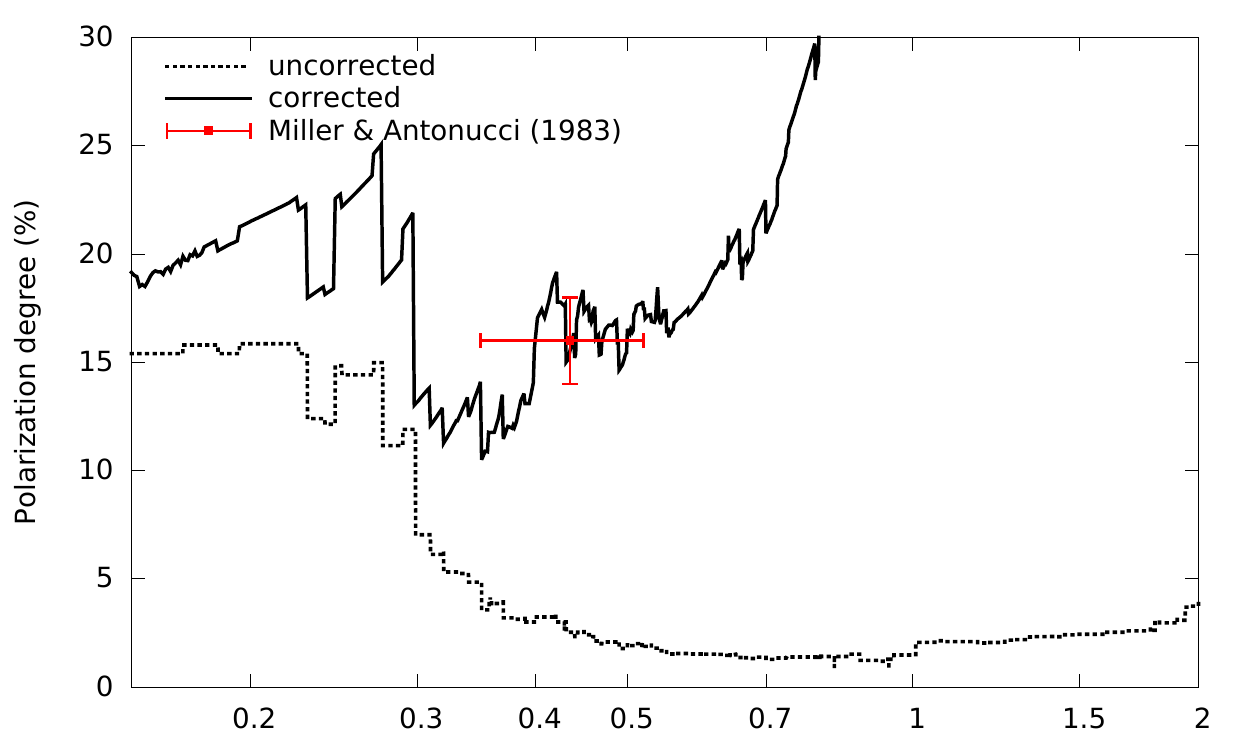}
    \caption{Corrected average continuum polarization of NGC~1068
	     using the fraction of diluting light (starlight 
	     plus dust emission components) in the observed continuum 
	     flux presented in Fig.~\ref{Fig:Fraction}.The previous 
	     estimation made by \citet{Miller1983} is shown in red.}
    \label{Fig:Corrected}
\end{figure}

In Fig.~\ref{Fig:Corrected}, we present the averaged combined polarization spectrum of NGC~1068 taken from Fig.~\ref{Fig:POL_spectrum} 
(dashed line), together with the linear continuum polarization corrected for 1) the host starlight dilution and 2) the dust
emission components, both the torus (i.e. reprocessed accretion disk emission) and the starburst dust emission, using 
the derived fraction of diluting light in the observed continuum flux of NGC~1068 (Fig.~\ref{Fig:Fraction}). We added the measure by \citet{Miller1983} 
in red and we cut the spectrum at 2$\mu$m since dust reemission is dominating at longer wavelengths. Our correction to the averaged compiled 
polarization spectrum of NGC~1068 reproduces the measurement by \citet{Miller1983} in the optical band. Our data indicates a median polarization of 
17 $\pm$ 3\% in the 3000 -- 7000\AA~ band while \citet{Miller1983} estimated a continuum polarization of 16 $\pm$ 2\% in the 
3500 -- 5200\AA~waveband. Our method is subject to higher polarization fluctuations due to the large range of observational apertures 
used but gives reliable estimation of the true scattered polarization of Seyfert-2 AGN. The sharp features at $\sim$ 0.22$\mu$m and 
$\sim$ 0.3$\mu$m are directly caused by the lack of polarimetric measurements in several consecutive bins (as it can be seen in the 
uncorrected spectrum in dashed line that sharply decreases at those wavelengths). It also appears that the continuum polarization increases 
in the ultraviolet due to dust scattering at distances larger than 1$"$ from the AGN core \citep{Antonucci1994}. This is in complete agreement 
with the discovery of \citet{Honig2013} and \citet{Asmus2016}, who have shown that optically thin dust in the polar outflows is responsible 
for much for the observed mid-infrared flux. The ultraviolet wavelength-dependent polarization signature we observe in Fig.~\ref{Fig:Corrected}
(for apertures greater than 1$"$) is naturally explained by scattering of light by dust particles along the polar direction. In the 
near-infrared band the asymptotic behavior of the dilution-corrected polarization spectrum is a consequence of the fact that the accretion disk 
continuum decreases relative to the stellar continuum, as wavelength increases. Nevertheless, we demonstrated that it is possible to build a 
reliable combined spectrum of the optical linear continuum polarization of AGN from published data despite the diluting action
of the host galaxy and dust emission components.

\subsection{Temporal evolution of the continuum polarization}
\label{data:Time_evolution}

The last aspect of our study is to test whether the polarization of NGC~1068 has significantly evolved in time. We know that the observed
polarization does not linearly depend on the amount of photons produced by the central engine. The chaotic light curve of AGN do not 
change the polarization state unless the powerful radiation field wipes out a fraction of the material in the vicinity of the black hole 
\citep[see, e.g.,][]{Lawrence1991,Jackson1990,Hill1996}. Even in this case, the amount of material removed by the radiation field must 
be significant to alter the polarimetric signature of the AGN in a detectable fashion \citep{Marin2016}. The dynamical timescales for 
changing the geometric arrangement of matter or magnetic fields at (sub-)parsec-scales around a 10$^8$ solar masses black hole is of 
the order of 5 -- 100 year \citep{Hopkins2012}. Only long-term monitoring of the polarization of a given AGN can probe such changes. 

\begin{figure}
    \includegraphics[trim = 0mm 0mm 0mm 0mm, clip, width=8.5cm]{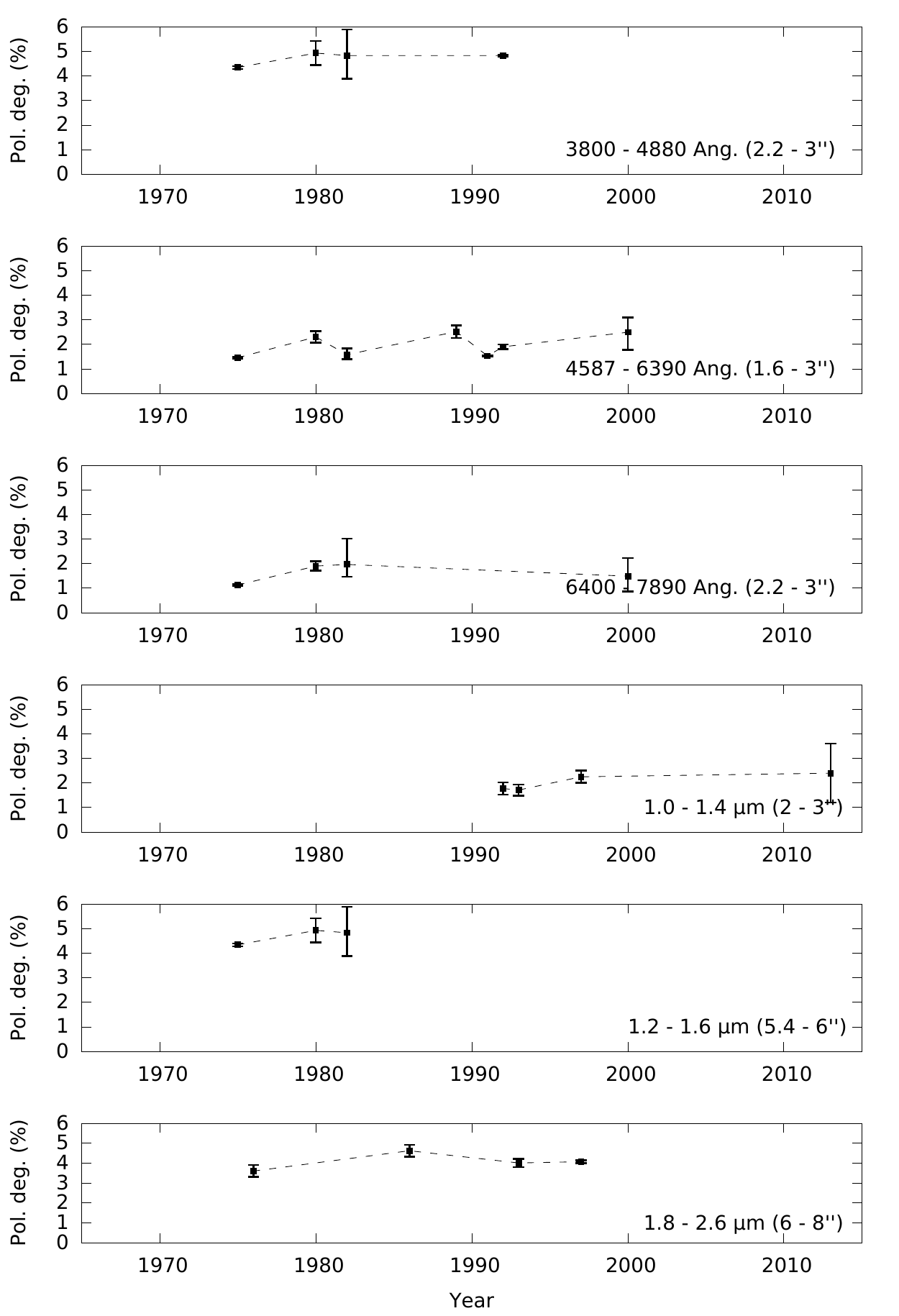}
    \caption{Time evolution of the polarization degree 
	     of NGC~1068 in several wavebands at a fixed
	     aperture. }
    \label{Fig:Time_variation}
\end{figure}

We present in Fig.~\ref{Fig:Time_variation} the time evolution of the continuum polarization of NGC~1068 in several wavebands
from the optical to the infrared domain. A similar aperture, indicated in each plot, was chosen when compiling the published data in 
order not to include systematic effects. There can still be (for the smallest apertures) a small impact of the seeing but this is
likely mitigated by the use of a range of apertures rather than a single aperture. As it can be seen, the polarization degree does 
not change by more than a fraction of a percentage point over several decades. If the error bars of the first and last polarimetric
points in the 3800 -- 4880\AA~and 1.8 -- 2.6$\mu$m figures are real, then there is a tentative statistically significant increase 
of polarization with time but the polarization angle (not shown) remains constant. This could coincide with the variability 
of the near-infrared nuclear flux of NGC~1068 observed between 1976 and 1994 by \citet{Glass1997}. From his study, it was not 
clear whether this event has been the response of an extended dusty region to a single outburst in the central engine or the 
result of a continuous change in its ultraviolet output, but it was shown in \citet{Marin2016} and \citet{Marin2017b}
that parsec-scale morphological changes in the geometry of the AGN would have varied the observed polarization degree by several 
percentage points and may be accompanied by a rotation of the polarization position angle. It is then safe to say that the global 
morphologic and magnetic geometry of NGC~1068 probably did not change since 1975. The 1965 -- 1975 period could not be investigated 
due to the large apertures used at that time, but it is unlikely that a strong change happened in less than a decade. Overall, we 
have proven that NGC~1068 remained remarkably constant in terms of polarimetry over the past 50 years.

\section{Multi-wavelength polarimetry of AGN: what is missing?}
\label{Discussion}
Reconstructing the polarization spectrum of an AGN is a difficult task since polarimetric data are easily contaminated by 
starlight, Galactic dust, starburst light and interstellar polarization. However, we have shown that the polarimetric spectrum 
of NGC~1068, despite the diversity of instruments and apertures, has distinctive wavelength-dependent signatures that can be related 
to the AGN and host galaxy physics. Impact of the Big Blue bump and infrared bump, electron, Mie and dichroic scattering, 
synchrotron emission and radio dilution by thermal electron are all naturally highlighted in polarization. This is the reason why 
polarimetry remains one of the best methods to explore the geometry and physics of unresolvable/obscured cosmic sources. Nevertheless
an important fraction of the polarized spectrum of NGC~1068 remains unexplored. 

The X-ray band is a territory where no polarimetric information has ever been recorded in the case of AGN. X-ray polarization 
is expected to arise from electron, dust and gas scattering, as well as magnetic processes \citep{McNamara2009}. Compton and 
inverse-Compton scattering are the dominant processes, similarly to Thomson and Mie scattering in the optical band, and the 
expected X-ray polarization can be evaluated (at first order) by looking at the visual polarization \citep{Marin2016}. In addition, 
since the X-ray continuum is produced close to the central supermassive black hole, special and general relativistic effects are 
expected to impact the observed polarization properties of light \citep[e.g.][]{Connors1977,Pineault1977,Connors1980}. Due to 
strong gravity effects, the polarization direction rotates along the photon null geodesics as the polarization vector is 
parallelly transported, resulting in specific polarimetric features. By observing the X-ray polarization of compact objects, 
it becomes possible to characterize the mass of the black hole, its spin, the composition of the accretion disk and the geometry
of the system \citep{Dovciak2004,Schnittman2009,Schnittman2010,Marin2017a,Marin2018}. 

The far-ultraviolet band offers unique insight into the physics of AGN that is still little known, in particular by probing 
ultraviolet-emitting and absorbing material arising from accretion disks, synchrotron emission in jet-dominated AGN and large-scale 
outflows. Some key signatures of accretion disks can be revealed only in polarized light, and with higher contrast at ultraviolet 
than at longer wavelengths \citep{Kishimoto2008}. Specifically, ultraviolet polarimetry can provide geometrical, chemical and 
thermodynamical measurements of accretion disks at unprecedented resolutions. By probing the ubiquitous magnetic fields, which 
are expected to align non-spherical small dust grains on the scales of the accretion disk to the extended torus, a future ultraviolet
polarimeter such as POLLUX on-board of LUVOIR \citep{Bolcar2016,Bouret2018} will be able to reveal the mechanisms structuring the 
multi-scale AGN medium. 

Far-infrared polarimetry remains an almost uncharted waveband despite the fact that observations do not suffer from seeing problem, 
allowing to deeply probe high extinction areas. Polarimetric measurements of ultra-luminous galaxies would allow one to address 
a variety of issues such as the nature of the emission that could be either non-thermal or originating from dust. By detecting 
solid state spectral features, far-infrared polarimetry should be able to discriminate between thermal and non-thermal emission 
in AGN, such as starburst galaxies \citep{Bressan2002,Andreani2003}. The large-scale  polarimetric signature most likely originates 
from extended magnetic fields that align dust grains. The associated dichroic absorption and emission mechanisms provide information 
on the geometry of the magnetic fields, which is essential to better understand accretion processes, disk formation and mass outflows
from stars to AGN \citep{Hough2003}.

Polarimetric imaging in the (sub)millimeter offers the possibility of identifying magnetic field configurations at unprecedented 
scales. Using ALMA, we can resolve the torus outer regions and study the polarized dust emission by dichroic emission and absorption
by aligned grains. In particular \citet{Aitken2002} have shown that sub-millimeter polarization can lead to strong constraints on 
the field configuration for a variety of torus models. Circular polarization, that is easily  acquired at those wavelengths, could 
also enlighten us on the jet structure of AGN \citep{Sazonov1969}. The POLAMI program \citep{Agudo2018,Agudo2018b,Thum2018} is using
the fact that Faraday conversion can convert linear polarization to circular polarization to probe the magnetized plasma of jet in a
large sample of radio-loud AGN. 

Finally AGN radio polarization measurements are possible as synchrotron emission is naturally highly polarized when magnetic 
field lines are ordered \citep{Westfold1959}. However, in the case of radio-quiet AGN polarization, the GHz and MHz bands remain
largely unexplored as Seyfert galaxies are generally reported to be unpolarized in the radio band, even in the optically thin regions 
\citep{Antonucci1993}. Yet those measurements have been achieved almost four decades ago and NGC~1068 appears to be (weakly) polarized at 
4.9~GHz. New observations with modern radio telescopes are necessary to explore with better sensitivity the emission and reprocessing 
mechanisms in the radio band. In particular the turnover frequency between dust polarized emission and synchrotron polarization could 
be detected by discovering the wavelength at which the polarization position angle rotates.

\section{Summary and conclusions}
\label{Conclusions}
We have gathered all the published polarimetric information on NGC~1068, the most observed Seyfert galaxy in terms of polarimetry.
We compiled the broadband 0.1 -- 100~$\mu$m, 4.9~GHz and 15~GHz continuum polarization of this archetypal type-2 AGN using data from more than 50 
years of observations, a premiere in the field. By doing so, we were able to detect all the expected transition regions in the 
polarized SED of NGC~1068: the Big Blue Bump (0.1 -- 0.7$\mu$m), the peak of starlight contribution ($\sim$1$\mu$m), the infrared
bump (1 -- 4$\mu$m), and the transition between electron scattering and polarized dust reemission (4 -- 5$\mu$m). Additional radio measurements 
also point toward another polarization mechanism linked with synchrotron emission, scattering and dilution by thermal electrons, 
highlighted by a rotation of the polarization position angle between the far-infrared and the radio bands. Despite the large 
variety of apertures and instruments used, the wavelength-dependent behavior of the polarization clearly revealed the wavebands where
processes are switching. This work allowed us to reconstruct the broadband polarized SED of NGC~1068 which strongly 
resembles to a typical type-1 SED seen in total flux. We analyzed the aperture effect on the measured continuum polarization of NGC~1068 
and found a large scale polarization component that can only be attributed to the host itself. This constant value for high apertures 
corresponds to the measured polarization degree of regular spiral galaxies, confirming previous observational results 
\citep{Scarrott1991,Simmons2000}. For apertures lower than 15$"$, the degree of polarization exponentially increases since the aperture 
gets smaller than the scattering region size, highlighting the presence of the narrow line region.

In order to carefully remove the contribution of starlight to the continuum polarization of our compiled spectrum, we reconstructed 
the global SED of NGC~1068 using nuclear fluxes from NED and a set of emissive components. By doing so, we were able to derive the 
fraction of starlight in the observed continuum flux of NGC~1068. We corrected our 0.15 -- 2~$\mu$m continuum polarization spectrum 
and found a wavelength-independent polarization level in the optical band, such as expected from theory and past studies. We also 
highlighted the impact of dust scattering in the ultraviolet and near-infrared bands for larger apertures (electron scattering is
dominating from the ultraviolet to the near-infrared at arc-second scale apertures). Finally, we checked whether the observed polarimetric 
signal of NGC~1068 varied through time. We demonstrated that this particular AGN did not undergo a major morphological or magnetic 
change over the past decades since its polarimetric signal remained constant (at similar apertures and in the same waveband).

In conclusion, we have created the very first broadband compiled polarization spectrum of an AGN exploiting more than 50 years of data.
The results are in strong agreements with the Unified Scheme of AGN despite the scarcity of data at several wavebands. Large class 
(30m) telescopes equipped with polarimeters are needed to pursue the study at deeper levels, both in spectroscopic and imaging modes.
Despite being restricted to 0.1 -- 100~$\mu$m, plus two points in the centimeter band, the work achieved in this paper can now be compared to the results of Monte Carlo radiative
transfer simulations in order to test different geometries of the reprocessing and emitting media \citep[e.g.,][]{Goosmann2007,Marin2012,Marin2015,Rojas2017,Grosset2017,Marin2018b}. 
This is an important step as, so far, any model-to-data comparison was only achieved in narrow bands. We also highlight the fact that 
our study could be expanded by many orders of magnitude by looking at the radio, millimeter, far-infrared, far-ultraviolet and X-ray polarization
of AGN using ALMA, HAWK+, or the forthcoming satellites IXPE \citep{Weisskopf2016}, eXTP \citep{Zhang2016} and LUVOIR \citep{Bolcar2016,Bouret2018}.

\section*{Acknowledgments}
This paper is a tribute to all the observers who dedicated their career to better understand AGN using polarimetry. I would 
like to thank (in alphabetical order) Beatriz Ag\'is Gonz\'alez, Robert Antonucci, Lucas Grosset, Damien Hutsem\'ekers, 
Enrique Lopez Rodriguez, Makoto Kishimoto, Chris Packham, Dominique Sluse, Yelena Stein and Bernd Vollmer for their numerous comments and suggestions that 
greatly improved this paper. Georges Roudnitski, from the Astronomicheskii Tsirkulyar journal, was of a great help to retrieve 
and translate old Russian papers. I also acknowledge the anonymous referee who helped to clarify and improve this paper.
Finally, the author would like to thank the Centre national d'\'etudes spatiales (CNES) who funded this project through to 
the post-doctoral grant ``Probing the geometry and physics of active galactic nuclei with ultraviolet and X-ray polarized 
radiative transfer''.

\label{lastpage}
\end{document}